\documentclass{article}
\usepackage{PdMaxCon25template}
\usepackage{times}
\usepackage{ifpdf}
\usepackage{soul}
\usepackage[english]{babel}

\def\papertitle{Preserving Russek's ``Summermood" Using Reality Check and a DeltaLab DL-4 Approximation}

\def\firstauthor{Jeremy Hyrkas} 
\def\secondauthor{Pablo Dodero Carrillo}
\def\thirdauthor{Teresa Díaz de Cossio Sánchez}

\newif\ifpdf
\ifx\pdfoutput\relax
\else
   \ifcase\pdfoutput
      \pdffalse
   \else
      \pdftrue
  \fi
\fi

\ifpdf 
  \usepackage[pdftex,
    pdftitle={\papertitle},
    pdfauthor={\firstauthor, \secondauthor, \thirdauthor},
    bookmarksnumbered, 
    pdfstartview=XYZ 
   ]{hyperref}

  \usepackage[pdftex]{graphicx}
  \graphicspath{{./figures/}}
  \DeclareGraphicsExtensions{.pdf,.jpeg,.png}

  \usepackage[figure,table]{hypcap}

\else 
  \usepackage[dvips,
    bookmarksnumbered, 
    pdfstartview=XYZ 
  ]{hyperref}  

  \usepackage[dvips]{epsfig,graphicx}
  \graphicspath{{./figures/}}
  \DeclareGraphicsExtensions{.eps}

  \usepackage[figure,table]{hypcap}
\fi

\hypersetup{
    colorlinks,%
    citecolor=black,%
    filecolor=black,%
    linkcolor=black,%
    urlcolor=black
}

\title{\papertitle}

 \threeauthors
   {0.5in}
   {\firstauthor} {\\ %
     {\tt \href{mailto:jhyrkas@ucsd.edu}{jhyrkas@ucsd.edu}}}
   {\secondauthor} {University of California San Diego \\ %
     {\tt \href{mailto:pdodero@ucsd.edu}{pdodero@ucsd.edu}}}
   {\thirdauthor} {\\ %
     {\tt \href{mailto:atdiazdec@ucsd.edu}{tdiazdec@ucsd.edu}}}

\begin{document}
\capstartfalse
\maketitle
\capstarttrue
\begin{abstract}
As a contribution towards ongoing efforts to maintain electroacoustic compositions for live performance, we present
a collection of Pure Data patches to preserve and perform Antonio Russek's piece ``Summermood" for bass flute and live electronics.
The piece, originally written for the DeltaLab DL-4 delay rack unit, contains score markings specific to the DL-4.
Here, we approximate the sound and unique functionality of the DL-4 in Pure Data, then refine our implementation to better match
the unit on which the piece was performed by comparing settings from the score to two official recordings of the piece.
The DL-4 emulation is integrated into a patch for live performance based on the Null Piece, and regression tested using the
Reality Check framework for Pure Data.
Using this library of patches, Summermood can be brought back into live rotation without the use of the now discontinued DL-4.
The patches will be continuously tested to ensure that the piece is playable across computer environments and as the Pure Data programming language is updated.
\end{abstract}

\section{Introduction}
\label{sec:intro}

The preservation of electroacoustic works is an important and difficult problem given that many pieces feature electronics that invariably go out of production or become obsolete.
Various approaches and initiatives have been put forth, and the restoration and preservation of important works of electronic music is ongoing work across many institutions.
Given the lineage of these institutions and the works they consider historically significant, many preserved works have an origin in Europe or North America, with works from other regions of the world perhaps underrepresented.
In particular, there is a need to preserve electroacoustic works from Mexico~\cite{Odgers2000}.
To contribute to the preservation of Mexican electroacoustic pieces, we focus here on the technical details of a recent restoration project focused on Antonio Russek's 1981 piece \emph{Summermood}.

\begin{figure}
    \centering
    \includegraphics[width=\linewidth]{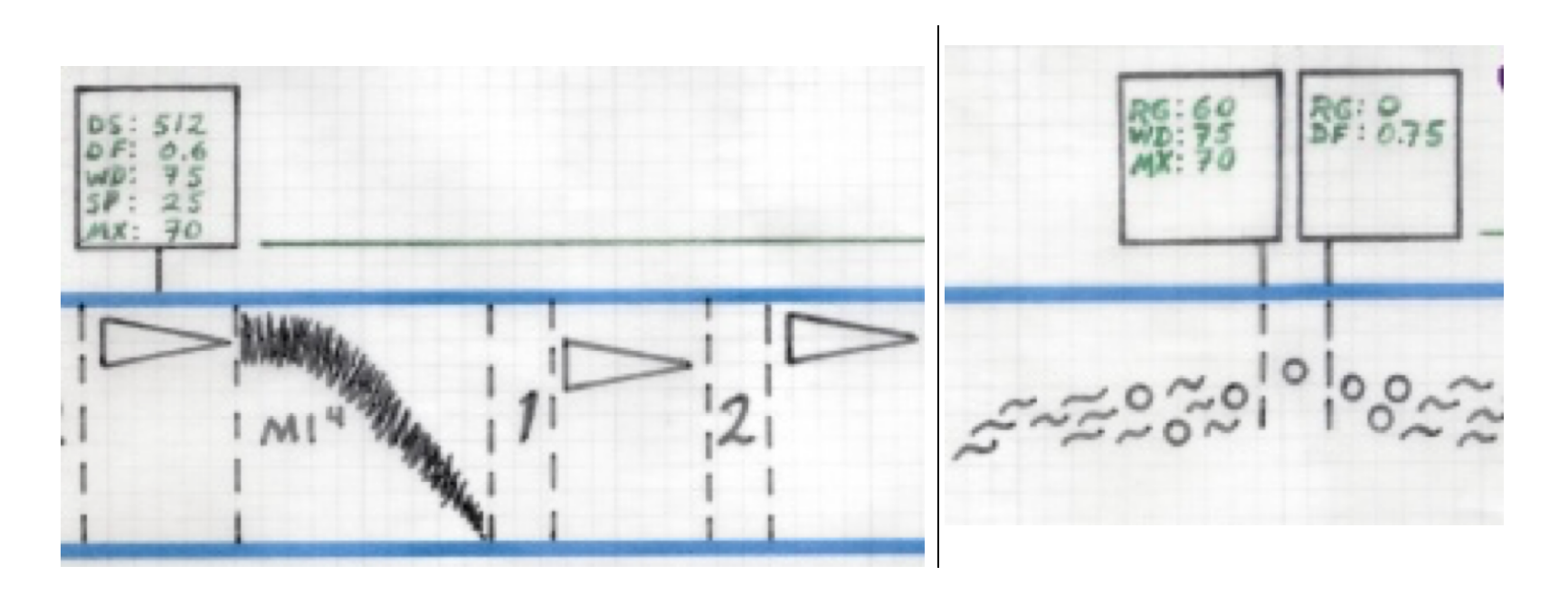}
    \caption{Two snippets from the \emph{Summermood} graphic score, including settings for the DL-4 in the upper margins (credit: Antonio Russek).}
    \label{fig:score}
\end{figure}

Russek composed Summermood for bass flute and live electronics.
The piece was written for and in collaboration with flutist Marielena Arizpe and used the DeltaLab DL-4 rack delay.
The score contains graphic notation for extended flute techniques, and delay settings that are specific to the DL-4's architecture and unique control set (see Figure \ref{fig:score}).
Summermood was performed solely by Aripe and only using Russek's personal DL-4 unit.
The original incarnation of DeltaLab closed in the early 1990s after the retirement of founder Richard DeFreitas, and the DL-4 was discontinued.
Arizpe similarly retired from public performance, and to our knowledge the piece has not been performed since the late 1980s.

The goal of this project is to create a working Pure Data simulation of the electronics used in Summermood so that it can be brought back into rotation as a live piece.
First, we emulate the DL-4 both in terms of its sonic quality and in terms of its particular control scheme.
Given the general lack of available DL-4 units, we cannot employ a typical approach to gear emulation where comparisons are made against a working unit.
Instead, we first implement an approximation of the DL-4 using its user manual as a guide.
We then refine the controls by comparing the score markings of Summermood against two recordings of the piece, creating a control set that closely approximates Russek's DL-4 and therefore the conditions under which the score and performances took place.
Finally, we integrate our DL-4 approximation into Miller Puckette's Null Piece and Reality Check frameworks, which allow the piece to be easily shared for live performances and to be continuously tested to confirm that it is still playable as Pure Data is updated.


\begin{figure*}[ht]
    \centering
    \includegraphics[width=\linewidth]{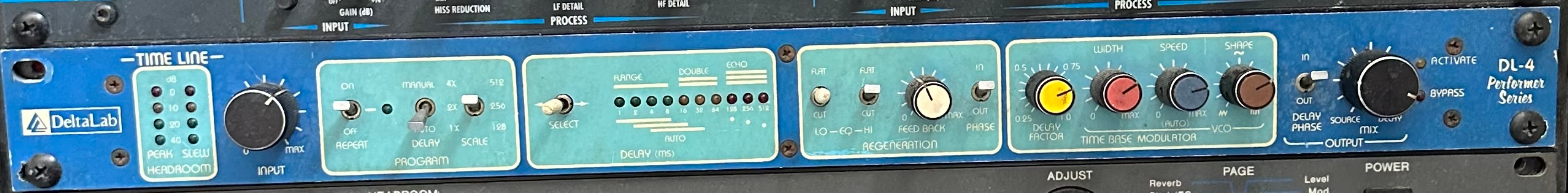}
    \caption{The faceplate of a DeltaLab DL-4 unit taken by the author.}
    \label{fig:dl4}
\end{figure*}

\section{Related work}
\label{sec:related}
Emulating discontinued musical gear and preserving electroacoustic compositions are separate problems that occasionally intertwine.
Some examples include Nicolas Collins' \emph{Pea Soup}~\cite{Collins2021}, originally written for the Countryman 968 Phase Shifter and Kaija Saariaho's \emph{NoaNoa}~\cite{Hodjati2013}, written for external reverb and pitching shifting units, both of which were later emulated in Max.
Morton Subotnick wrote pieces for \emph{ghost boxes}, custom-built electronic devices which he later emulated using Max~\cite{Hanson2011}.
Tom Erbe worked with James Tenney to port his piece \emph{For Ann (rising)}, originally created using a Lafayette oscillator and tape, to an implementation using Csound~\cite{Dennehy2008}.
Other examples likely exist, mostly as piecemeal efforts without a central preservation platform or framework.

In an effort to create a more standardized and maintainable approach to electronic music preservation, Miller Puckette proposed the Null Piece~\cite{Puckette2021} and Reality Check~\cite{Puckette2024} frameworks, built using and on top of the Pure Data programming environment.
The Null Piece is a starting point for electroacoustic pieces.
It consists of a skeleton patch and a library to assist with typical components of a piece for live electronics, which can include: audio inputs, gain control, live audio processing, time-dependent presets and parameter control, and audio outputs.
A piece of music built in the Null Piece format can be maintained and shared so that anyone with Pure Data and any necessary dependencies can perform the piece.

Reality Check allows for a Pure Data patch to be run in a virtual environment to check that it still functions as intended.
If the patch, processing a predefined audio input with specified parameter changes at given timestamps, produces an audio file
virtually identical to a predefined output, 
then the patch behaves as expected given the current computer environment and Pure Data version.
The Null Piece and Reality Check frameworks dovetail into a straightforward mechanism for preservation:
the performance patch built using the Null Piece design is tested using Reality Check, and a successful test indicates that the piece is still playable using the performance patch. 

\section{The DeltaLab DL-4 and an emulation in Pure Data}
\label{sec:dl4}

The DL-4 is one of several delay rack models built by DeltaLab in the 1980s, inspired by their Effectron series~\cite{Harris1984}.
The series was known for its full-bandwidth delay section and long maximum delay time compared to contemporary analog delay effects.
While digital audio is commonly encoded using Pulse Code Modulation (PCM), the DeltaLab units encoded audio using Adaptive Delta Modulation (ADM)~\cite{Abate1967}.
The faceplate of a DL-4 is shown in Figure \ref{fig:dl4}.

\subsection{Delay time control}
DeltaLab delays featured a unique approach to delay time control.
A base delay time was chosen using a selector mechanism; delay times increase in powers of two, and often ranged from as little as 1 ms (for comb filtering) up to 512 or 1024 ms (for long digital repeats).
A continuous \emph{Delay Factor} (DF) knob controlled the clock rate of the unit, effectively acting as a multiplier on the base delay time.
For example, a base delay time of 256 ms with a DF of 0.6 results in a delay time of $256 \cdot 0.6 = 153.6$ ms.
On the DL-4, the Delay Factor control ranges from 0.25 to 1.0 times the base delay time.
This delay time control mechanism is reflected in the Summermood score; see Figure \ref{fig:score}, where the markings for \emph{DS} indicate the base delay and \emph{DF} mark the position of the DF knob.

The DL-4 features a low-frequency oscillator (LFO) that interacts with the DF settings.
The LFO speed ranges from 0--10 Hz (marked \emph{SP} in the Summermood score), and the shape control sweeps the wave shape between a triangle, sinusoid, and square wave.
According to the user manual, when the LFO width (\emph{WD} in the score) is turned to 100\%, the DF is disregarded and the base delay time is multiplied by the LFO sweeping between the values of 0.25 and 1.0.
When the width control is less than 100\%, the DF is added to the LFO wave to center the delay time sweep.

\begin{figure*}
    \centering
    \includegraphics[width=\linewidth]{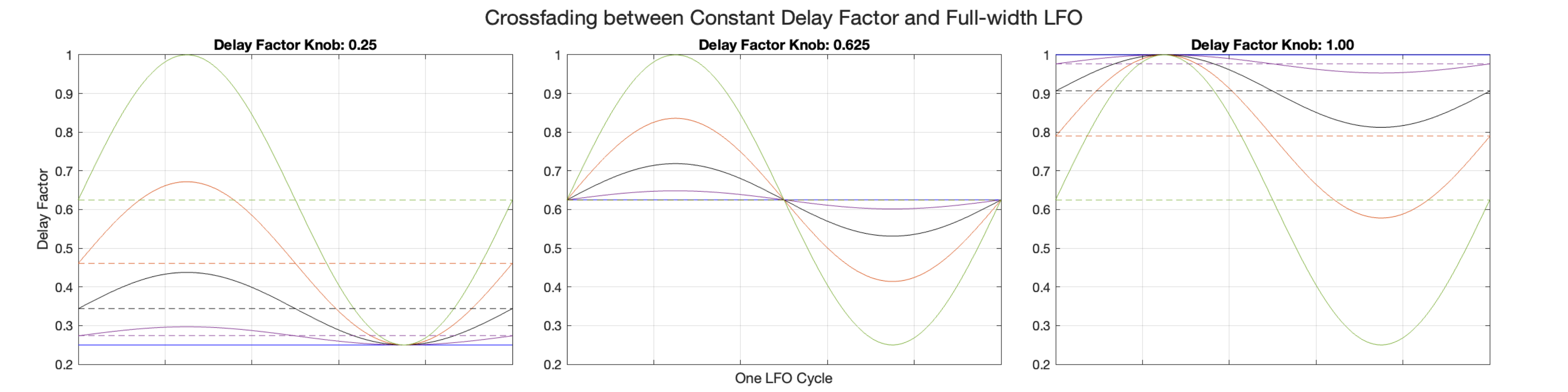}
    \caption{The estimated effect of the LFO Width knob.
    A linear crossfade between the constant Delay Factor setting and a constant waveform between 0.25 and 1.0 is used, causing the mean of the LFO (dashed lines) to drift towards the center position of 0.625 as the Width knob is increased.}
    \label{fig:lfo}
\end{figure*}

The addition of DF and LFO is ambiguously defined, as various parameter settings lead to a waveform whose amplitude exceeds 1.
Because this waveform acts as a multiplier on the delay time, the resulting delay length could exceed the unit's stated maximum 512 ms.
We attempt to mitigate this ambiguity by using the position of the LFO width knob to crossfade between the constant DF and the LFO sweeping between 0.25 and 1.
In this implementation, the DF is constant when the width knob is set to 0, and the mean of the modulating waveform drifts towards the center of the DF range as the width knob is increased.
Figure \ref{fig:lfo} depicts the effect of sweeping the width knob when the DF is set to its minimum, middle, and maximum positions.
This implementation ensures that the maximum and minimum delay times are always within the stated range of the DL-4, although it is unclear if this crossfade is the implementation used in a real DL-4. 

\subsection{Other user controls and signal path}

The DL-4 contains various other user parameters, most of which are standard to digital delays.
A feedback knob controls the gain of the signal fed from the delay line back into itself.
At its maximum, the amount of feedback is less than 100\%; this control is marked \emph{RG} in the Summermood score.
The feedback loop is analog and goes through a high- and low-pass filter, tuned to 16 Hz and 15 kHz with switches to change the cutoff frequencies to 150 Hz and 3.3 kHz, respectively.
The signal fed back into the delay can have its phase reversed, and the overall output of the delay line can also be phase reversed.
The dry-wet control is a linear mix (\emph{MX} in the Summermood score) of the input signal and the delay-line signal.
While most controls on the DL-4 are linear, logarithmic potentiometers are used for the LFO width and speed\footnote{Special thanks to Tom Erbe for an initial circuit analysis.}.

\subsection{Pure Data Implementation details}
\begin{figure}
    \centering
    \includegraphics[width=\linewidth]{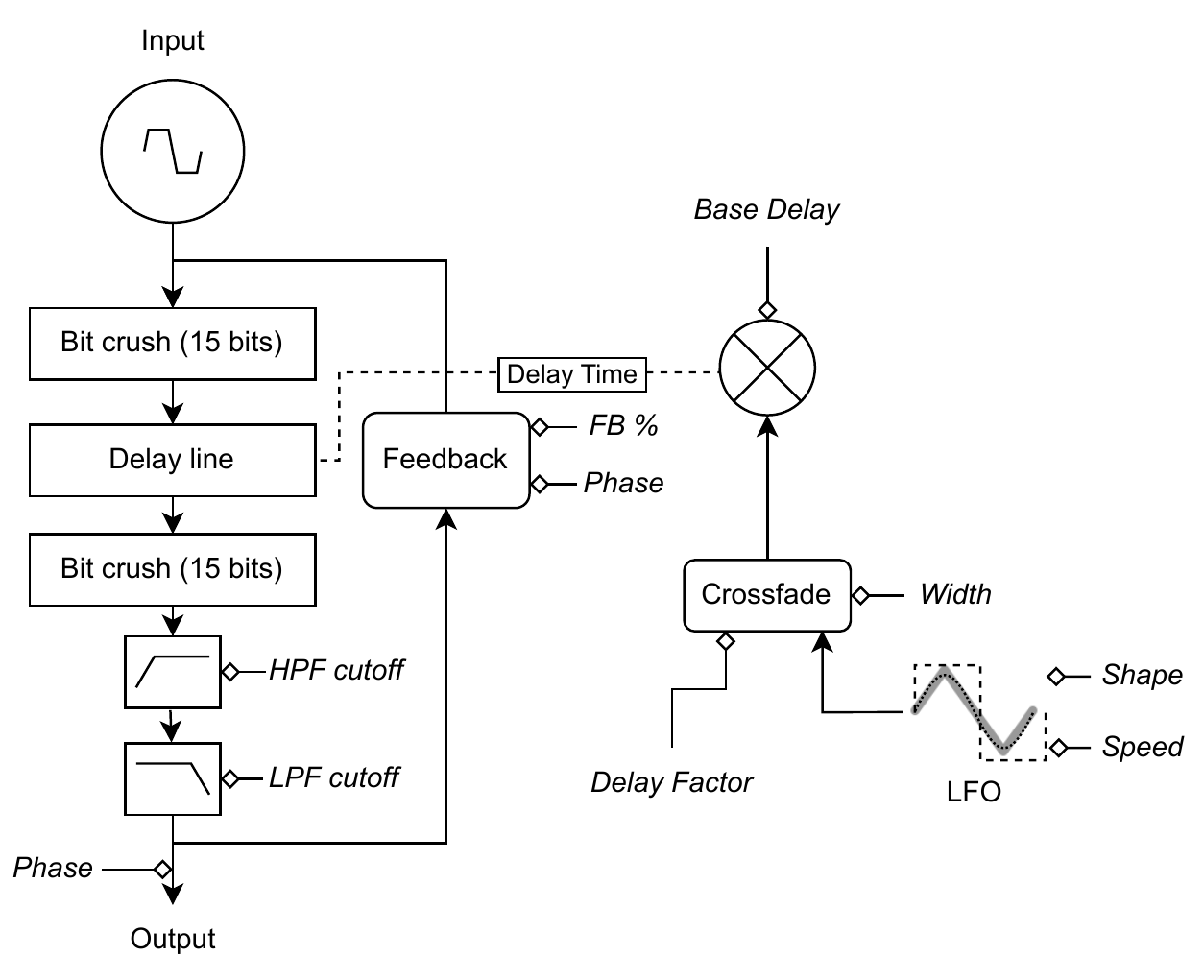}
    \caption{Signal path for the delay section of the DL-4 emulation. User parameters are indicated in italics and by lines with a diamond connector.}
    \label{fig:dl4_flow}
\end{figure}

We approximate the delay portion of the DL-4 using standard objects in Pure Data, with matching user-controlled parameters.
The signal path is shown in Figure \ref{fig:dl4_flow}, and the simulated faceplate of this implementation is shown later on the main patch of the Summermood Null Piece (see Figure \ref{fig:null_piece}).
While closer emulation may be achieved using external packages or custom externals (particularly to mimic the sonic effects of encoding using ADM), keeping the implementation in vanilla Pure Data more easily allows the Summermood piece to be shared and tested, as we will discuss in Section \ref{sec:null}.

The user manual states that the analog-to-digital conversion of the DL-4 has over a 90 dB signal-to-noise ratio (SNR).
In PCM signals (like those in Pure Data), this would correspond to a bit depth of around 15 bits.
Although DeltaLab units use ADM, we use a bit crusher set to 15 bits before the delay write and after the delay read to simulate the digital noise that might be present in a DL-4. 
The second bit crush is necessary because the feedback path in the DL-4 is analog.

The DL-4 features a handful of unique controls that are not currently implemented in our emulation and not used to perform Summermood, but may nevertheless be of interest in future revisions.
A \emph{Repeat} switch disconnects the input signal from the delay line and digitally recirculates the contents of the delay line.
Adjusting the DF controls the clock speed, causing a shift to the pitch and speed of the recirculated signal.
The \emph{Auto} delay function sweeps through the base delay using the LFO speed setting, switching between harmonically and rhythmically delays due to the factor-of-two relationship between base delay times.
Finally, the delay time of the DL-4 could be dynamically controlled with the use of an external controller.

\section{Refining the DL-4 emulation using Summermood}
\label{sec:summermood}

After our initial implementation of the basic functionalities of the DL-4, we compared the settings from the Summermood score against two recordings of the piece as performed by Arizpe using Russek's DL-4 unit.
In the process, several discrepancies between the recorded and expected signals were found, particularly with regards to the delay time, feedback amount, and the use of the LFO to modulate the delay time.
We refined the controls of our emulation to scale to those indicated in the score.
This refinement, while perhaps straying from the theoretical implementation according to the DL-4 manual, ultimately allows us to share a Summermood Pure Data patch that matches the official score and more closely recreates the conditions under which it was originally performed.

There are two official recordings of Summermood.
In the more prominently available version, the bass flute signal and delay unit signal are more or less hard panned to the left and right before an external reverb is applied.
Using this recording, we used careful listening to compare the left channel passed through our Pure Data DL-4 approximation to the right channel of the recording.
These signals are mostly comparable, with the exception that we must pass the flute with reverb applied into our emulation, whereas in the recording the flute is passed into the DL-4 and then into the reverb.
While we use this version primarily to test our patch, a second recording is used to refine our estimated settings, as this version is preferred by the composer.

\subsection{Feedback and Delay Factor range refinement}
In performances of Summermood, delay parameters are mostly static, with changes applied at once and no parameters swept or time-varied.
Our main parameters of interest are the Delay Factor knob position, the feedback knob position, and the LFO width and speed settings.

In general, feedback was tuned by counting the number of audible repeats.
This approach is accurate up to the point where repeats decay below the level of the reverb.
One section of Summermood utilizes a very short delay time and high feedback to create a resonant comb filter that is activated by impulsive sounds from the bass flute.
In this case, the true digital feedback was estimated by audibly matching the resonance and decay time of the delay.

\begin{figure}
    \centering
    \includegraphics[width=\linewidth]{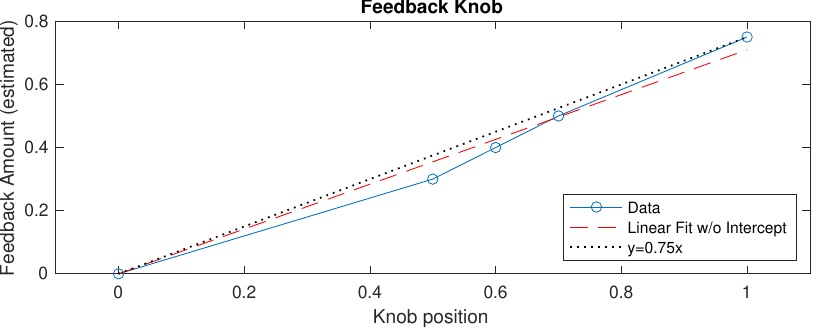}
    \caption{Regeneration: score settings versus estimated feedback.
    The basic map $y=0.75x$ map is a good fit.}
    \label{fig:regen}
\end{figure}

Using this method, we created a small data set of six score settings and their estimated ``true" feedback settings (i.e., the multiplicative gain in the feedback loop of a digital delay).
These data points were fit with a linear regression. 
This mapping is shown in Figure \ref{fig:regen}.
In our final version, we use the simple linear map $y=0.75x$, which is similar to the linear regression model and allows for a feedback gain of 0.75 which is necessary in the aforementioned comb filtering section of the piece.
The DL-4 manual states that the feedback knob cannot reach 100\% feedback, and a circuit analysis shows that the knob has an adjustable trim.
Given that the same unit was used for all official recordings and performances, it seems likely that this trim was set so that the unit allowed for a maximum of around 75\% feedback.

We also collected six data points mapping DF score settings and estimated true delay times.
Delay times were derived by examining repeat times in the time domain waveforms of the recorded performances, with ``true" Delay Factors derived as the estimated delay time divided by the base delay time indicated on the score.
In the section of the piece featuring the comb filter, the resonant frequency of the filter was found to be roughly 45 Hz, indicating a delay time of 22 ms.
This resonant frequency was consistent between both recordings of Summermood.

\begin{figure}
    \centering
    \includegraphics[width=\linewidth]{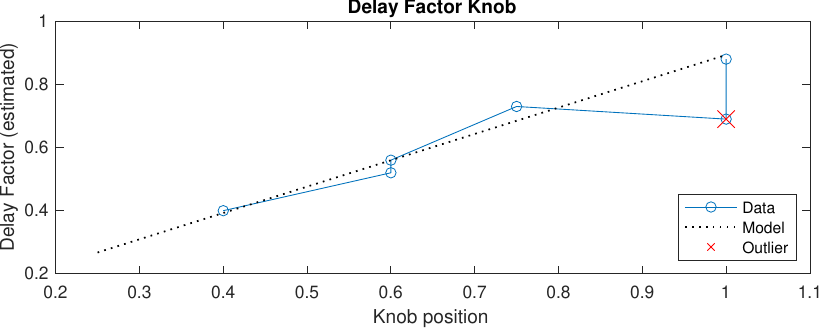}
    \caption{Delay factor: score settings versus estimated percent of base delay time.
    A reasonable map is found, but only by excluding an outlier.}
    \label{fig:df}
\end{figure}

As before, the data points were fit using linear regression.
We visually determined that the DF setting for the comb filtering portion of the piece was an outlier and excluded it from the regression.
The linear model maps the DF score settings to a true DF using $y = 0.05899 + 0.83434x$.
Because the DF knob sweeps from 0.25 to 1, the model reduces the actual range to roughly 0.27 through 0.89.
The data and map are shown in Figure \ref{fig:df}.
This mapping could be explained by a dirty or otherwise inaccurate potentiometer in the DL-4 unit used in the recordings.

We use this map in our emulation, which reduces the range of the DF control but ultimately allows our patch to more closely resemble the settings from the original score while still allowing parameters to be tweaked to taste in a live performance.
The sole exception is the outlier in Figure \ref{fig:df}; the score marking indicates a DF setting of 1.0 but our piece uses a value of 0.75 to ensure that the resonant frequency of the comb filter matches that of the recordings.

\subsection{The mystery of the LFO}
The Summermood score indicates settings for LFO width and speed throughout the piece.
Based on our understanding of the DL-4 manual, the settings indicated in the score would cause delay times sweeps of 10--20 ms (or greater).
Delay time changes of this magnitude cause a pitch shifting or Doppler effect in the delayed signal~\cite{Smith2002}.
We confirmed this behavior in videos of the DL-4 in use available online and tests of our own DeltaLab Effectron Jr. delay unit, which has similar LFO controls to the DL-4.
However, both recordings of Summermood do not feature pitch shifting and, in fact, exhibit remarkably consistent delay times, indicating that no modulation of delay time is used.
We are actively working with the composer to discover the source of this discrepancy.
For now, the LFO settings in the score are discarded in our live performance patch to more closely match Russek's vision of the piece.

\subsection{Spectral comparison of implementation}

\begin{figure*}
    \centering
    \includegraphics[width=0.9\linewidth]{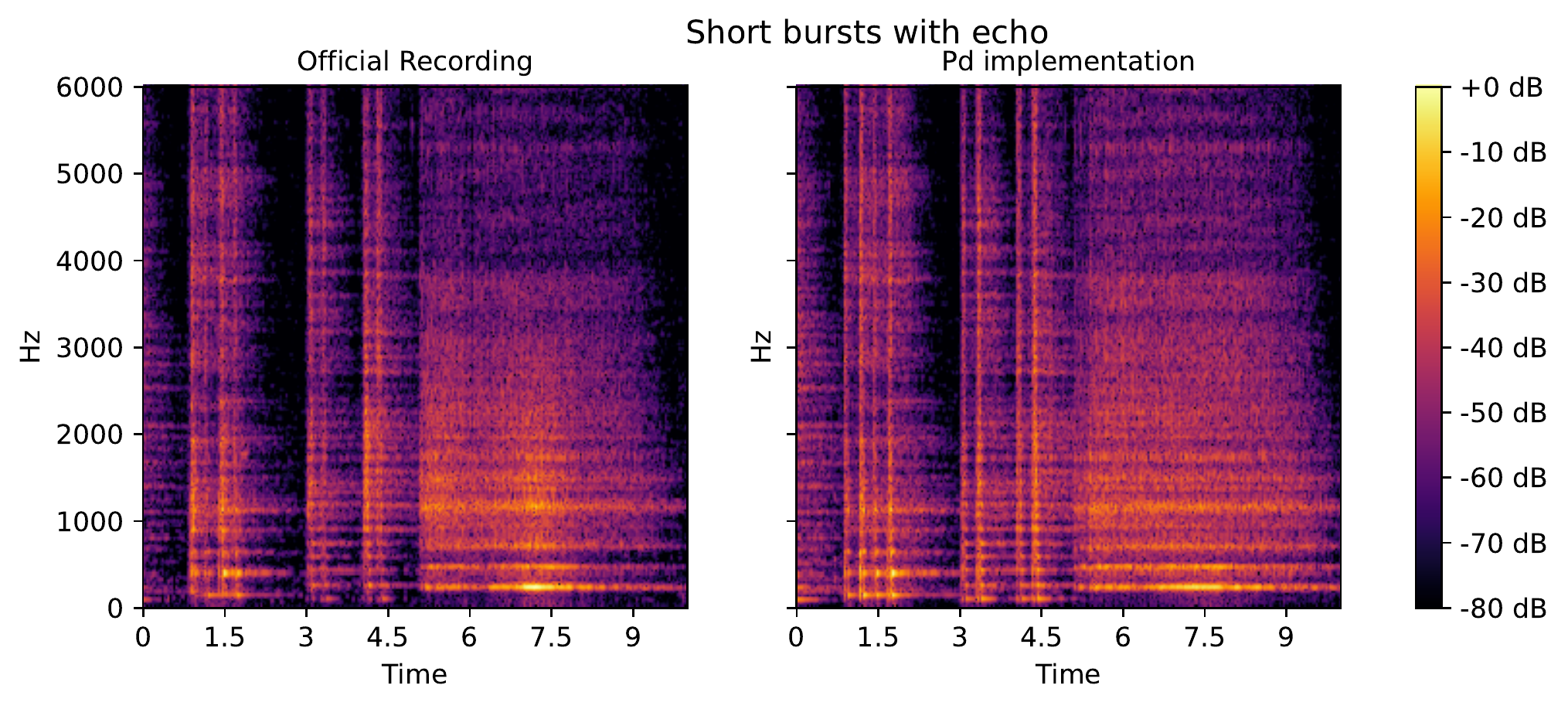}
    \caption{A section of Summermood (roughly 00m:15s) when the flutist plays short bursts while the DL-4 is set to a short echo.}
    \label{fig:bursts}
\end{figure*}

\begin{figure*}
    \centering
    \includegraphics[width=0.9\linewidth]{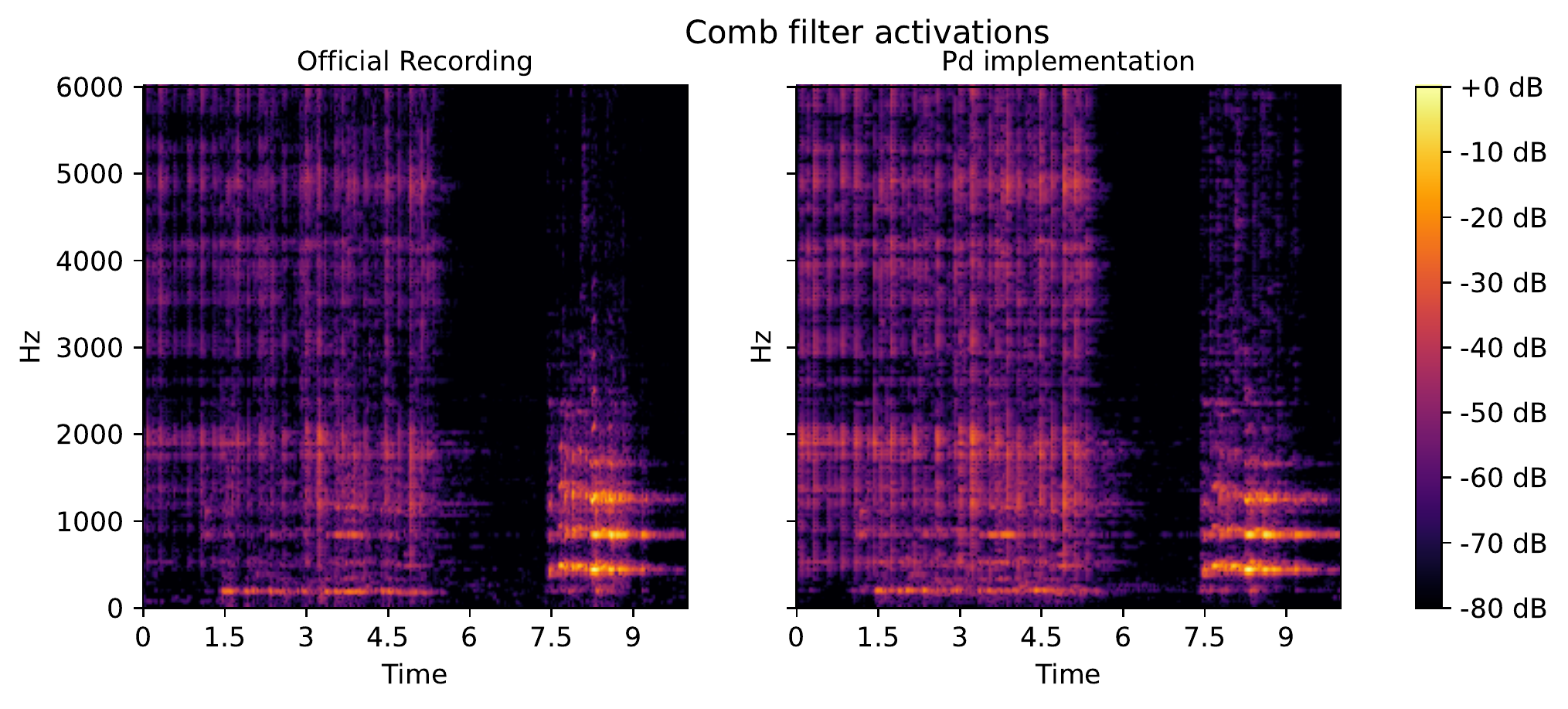}
    \caption{A section of Summermood (roughly 03m:07s) when the flutist plays key clicks to active the DL-4 set as a resonant comb filter.}
    \label{fig:combs}
\end{figure*}

\begin{figure*}
    \centering
    \includegraphics[width=0.9\linewidth]{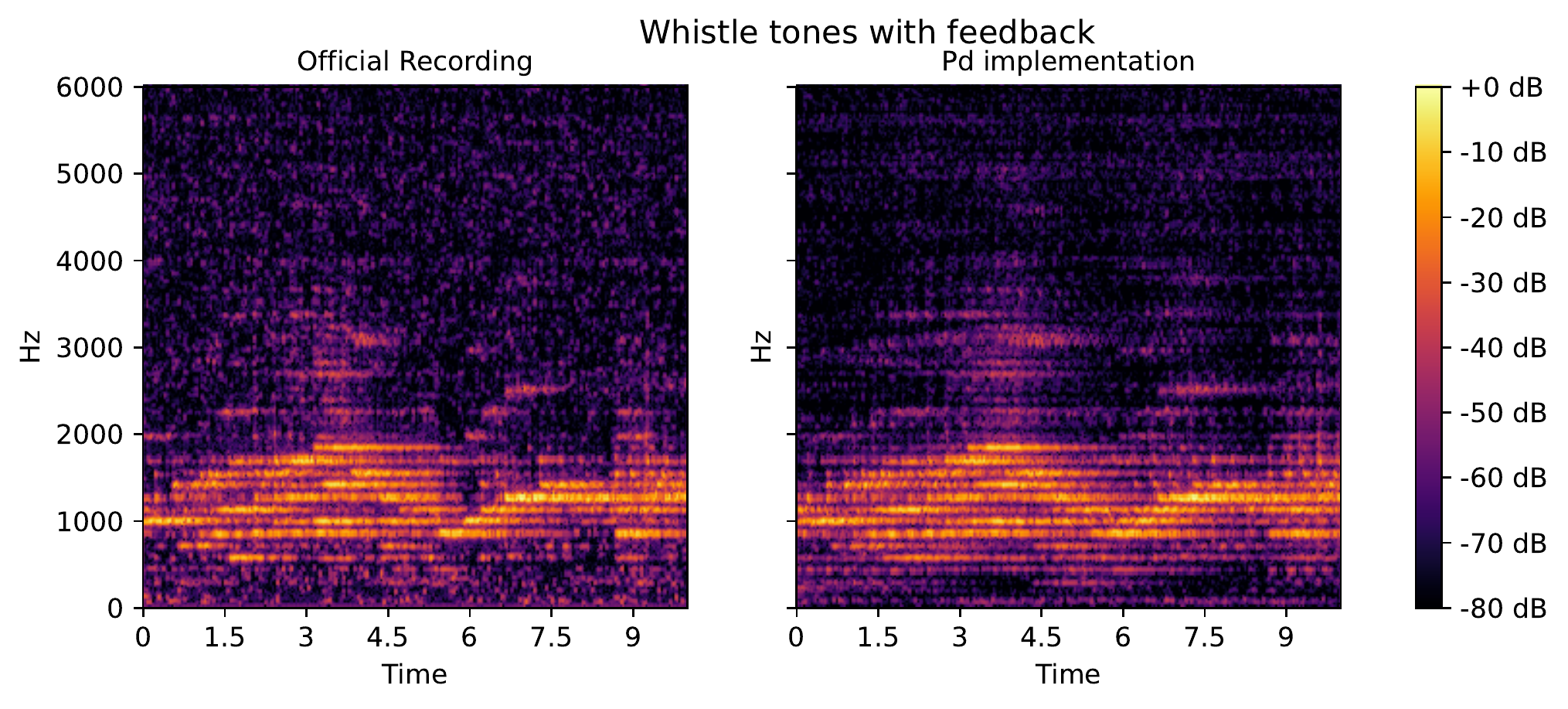}
    \caption{A section of Summermood (roughly 04m:18s) when the flutist plays whistle tones while the DL-4 is set to a short echo with feedback.}
    \label{fig:whistles}
\end{figure*}

Figures~\ref{fig:bursts}--\ref{fig:whistles} show comparisons between the right channel of the official recording of Summermood and the left channel run through the DL-4 emulation at three key moments of the composition.
In these examples, the DL-4 is configured as a short echo, a resonant comb filter, and an echo with feedback for atmospheric effect.
Although most of the spectral information is determined by the flute signal, the spectrograms reveal similar echo patterns in Figure~\ref{fig:bursts} and visible harmonics of the resonant comb filter in Figure~\ref{fig:combs}.
The Pure Data signals have a somewhat higher SNR than the official recording, but this is likely due to slightly different wet-dry settings between the official recording and the patch settings used for these examples.

\section{Performing and Preserving Summermood using the Null Piece and Reality Check}
\label{sec:null}

\begin{figure*}
    \centering
    \includegraphics[width=0.66\linewidth]{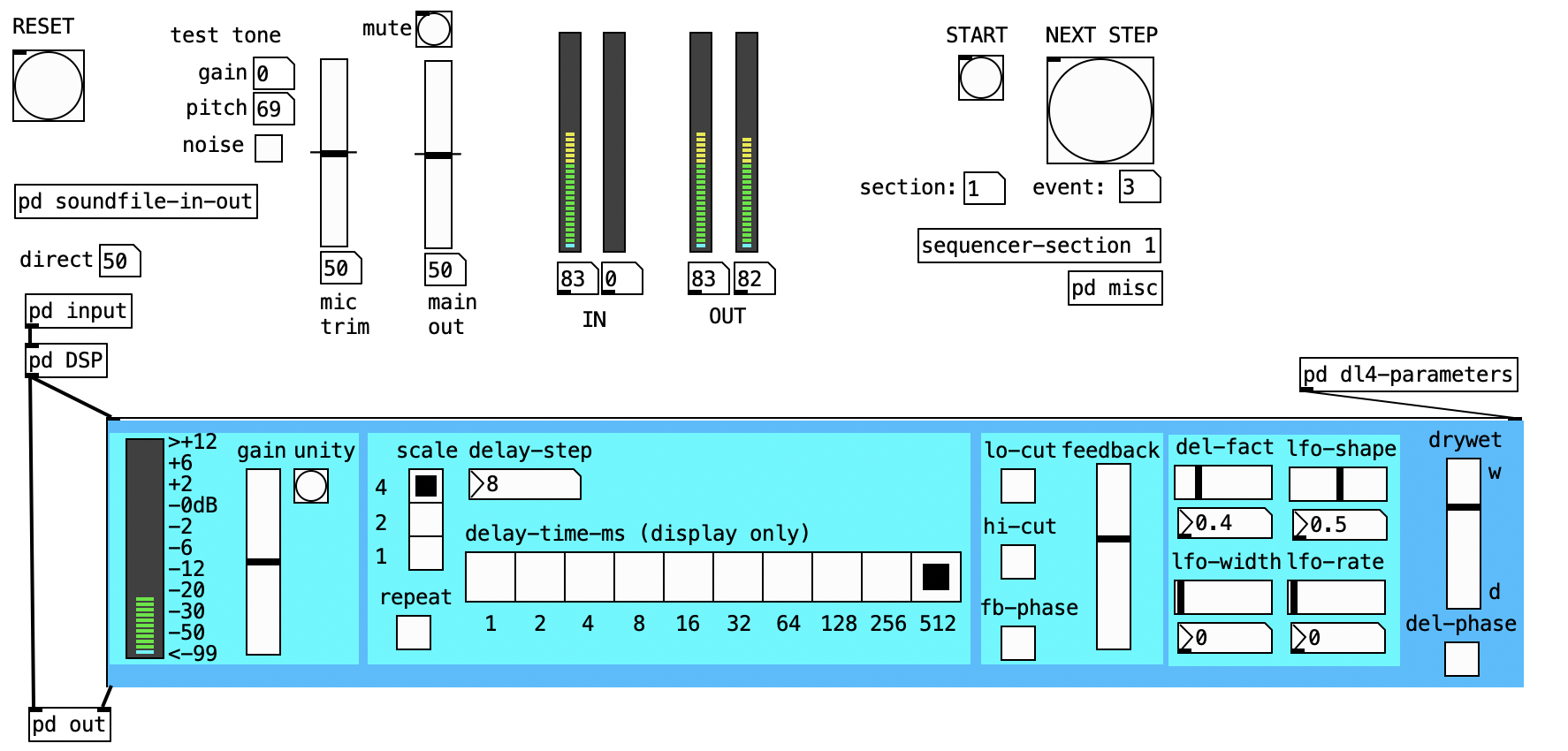}
    \caption{The Summermood Null Piece patch.}
    \label{fig:null_piece}
\end{figure*}

The final component of this project is a portable and maintainable Pure Data implementation of Summermood for live performance.
We integrated our DL-4 emulation into a skeleton patch based on Puckette's Null Piece~\cite{Puckette2021}.
The Null Piece library comes with two potential starting patches, and we use the patch ``1.pd" for its inclusion of a programmable sequencer using Pure Data's qlist.
We scripted an event list where each sequencer step corresponds to a setting from the Summermood score.
To perform the piece, the electronics operator should follow the score and advance to the next step at the appropriate time, adjusting the delay settings by hand if desired.
This patch can be modified so that steps are advanced using an external controller, which could be used by a flutist to play the piece solo.

Figure \ref{fig:null_piece} shows the main patch of our implementation.
In a typical Null Piece, the electronic processing takes place in the \emph{pd dsp} subpatch.
However, our implementation places the DL-4 emulation in the main patch so that the operator can interact with the parameters as they would using a physical DL-4.
Settings from the qlist are routed using the \emph{pd dl4-parameters} subpatch, and the emulation internally updates the GUI and internal DSP.

Included with this library is an initial test script that runs the Summermood patch against the left channel of the official Summermood recording with time-stamped sequencer changes using the Reality Check~\cite{Puckette2024} framework.
The timing is currently based on this performance and can be used to ensure that the patch creates the expected audio output with each revision.
We are in the process of working with Russek to create a new definitive recording of the piece.
This recording will be created using our patch and regression checked using Reality Check.
When this work is complete, we will be able to confirm that the Summermood patch is still playable and still creates the canonical version across operating systems and Pure Data revisions.

\section{Discussion}
\label{sec:discussion}
Work is ongoing towards creating the finalized Summermood patches and piece.
A preliminary performance is planned, which will act both as a live proof-of-concept and the first official Summermood performance since the 1980s.
The patch, score and test script will be updated as needed until Russek and Arizpe give their final approval.

While the main drive of this work is to preserve Summermood, there are opportunities to improve the DL-4 emulation by including more of the original functionality and testing the sonic qualities of the emulation against a working unit.
The DL-4 implementation associated with Summermood will likely diverge from these updates to maintain its portability, particularly if future DL-4 revisions include the use of custom Pure Data externals.

This paper reflects one aspect of the multifaceted process of electronic music preservation.
Future manuscripts are planned documenting the history of Summermood, Russek's role in Mexican electroacoustic music, Arizpe's influence as a flutist and the impact of electronic music from Latin America and its place in the musical canon.

\begin{acknowledgments}
The authors thank Antonio Russek, Marielena Arizpe, Miller Puckette and Tom Erbe for their invaluable assistance. 
\end{acknowledgments} 

\bibliography{main}

\end{document}